\documentclass[aps,twocolumn,prb,tightenlines,floatfix,showpacs,nofootinbib]{revtex4}
\usepackage[dvips]{graphicx}
\usepackage[english]{babel}
\usepackage{amsmath}
\usepackage{amssymb}
\usepackage{bm}
\usepackage{color}

\newcommand{\be}{\begin{eqnarray}}
\newcommand{\ee}{\end{eqnarray}}
\newcommand{\p}{\partial}

\begin{document}

\title{Intrinsic Anomalous Hall Effect in Magneto-Chiral States}
\author{Yan He}
\affiliation{College of Physical Science and Technology,
Sichuan University, Chengdu, Sichuan 610064, China}
\author{P. A. Lee}
\affiliation{Department of Physics, Massachusetts Institute of Technology, Cambridge, MA  02139}
\author{C. M. Varma}
\affiliation{Department of Physics and Astronomy, University of California, Riverside, CA 92521}

\begin{abstract}
We show that a finite Hall effect in zero applied magnetic field occurs for partially filled bands in certain time-reversal violating states with zero net flux per unit-cell. These states are the Magneto-chiral states with parameters in the effective one-particle Hamiltonian such that they {\it do not} satisfy the Haldane-type constraints for topological electronic states. The results extend an earlier discussion of the Kerr effect observed in the cuprates but may be applicable to other experimental situations.
\end{abstract}

\maketitle

\section{Introduction}
\begin{figure}
\includegraphics[width=0.3\textwidth]{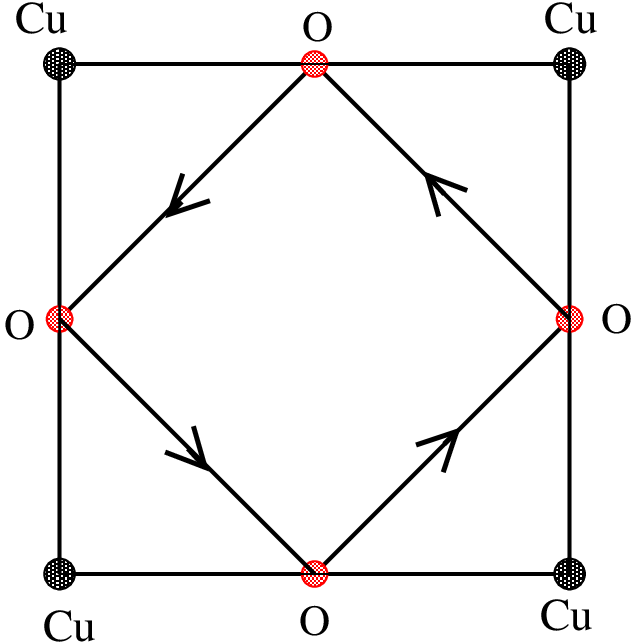}
\caption{The loop-current pattern of the Magneto-chiral or anomalous Hall effect state in Cuprates.}
\label{Fig-AHE}
\end{figure}

A Faraday rotation of the polarization of photons in transmission or a Kerr effect in reflection must accompany a state which has a non-zero off-diagonal conductivity $\sigma_{xy}$ \cite{LLP}. A Kerr effect with unusual properties has been discovered in underdoped cuprates \cite{kerr-expt} below a temperature which in two families of cuprates coincides with the so-called pseudogap temperature $T^*$ deduced through several transport and thermodynamic measurements \cite{pseudogap}.
Within some uncertainty, this is also the temperature below which evidence for a state with broken time-reversal has been presented in these compounds by polarized neutron scattering \cite{neutrons}. The symmetry of the state is consistent with that of  a magneto-electric (ME) state \cite{theory}. The ME state has no Berry phase thus cannot have a finite $\sigma_{xy}$ \cite{he-moore-v}, i.e. a Kerr effect in zero applied magnetic field. But as discussed in a recent paper \cite{aji-he-cmv}, a state with an anomalous Hall effect (AHE) can be induced in the ME state in the presence of perturbations which independently break the reflection symmetries of the lattice.  The induced state has the current pattern in a unit-cell exhibited in
Fig. (\ref{Fig-AHE}) and has been called a Magneto-Chiral (MC) state \cite{aji-he-cmv}. The discussion in
Ref. (\onlinecite{aji-he-cmv}) was for a three band model with parameters such that they are an extension of the topological state discussed by Haldane \cite{Haldane} for a two band model. As explained below for the three band model and as was already
done for a two band model by Haldane, topological AHE requires that a monopole-like singularity in the parameter space is enclosed by the torus of Brillouin zone. When this requirement is met, the integration of Berry curvature on the whole Brillouin zone is a quantized topological invariant. For later convenience, we call this requirement as ``Haldane condition".
The principal purpose of this paper is to show that non-quantized AHE occurs for magneto-chiral states in partially filled bands without satisfying the Haldane condition.

The general physical point we wish to make is that in a time-reversal violating state a current is carried by each eigenstate ${\bf k}$ of any of the bands. In the ME state, which besides violating time-reversal are also odd under inversion with the product of time-reversal and inversion conserved, an eigenstate ${\bf -k}$ carries current in the opposite direction to the state ${\bf k}$ at every point in space. So there is no net AHE. On the other hand in a MC state, ${\bf k}$ and ${\bf - k}$ carry current in the same direction. When the Haldane condition is not satisfied for the MC state, the sum of current over the complete Brillouin zone for any band is 0. But then there must exist a finite $\sigma_{xy}$ for a partially filled band.

Our result can be understood as a special case of the intrinsic anomalous Hall effect,\cite{11} where the Hall conductivity is written as
\begin{equation}
\sigma_{xy} = {e^2 \over \hbar}\sum_n \int {d\bm k \over (2\pi)^{\bm d}} F_z(\bm k) f(\varepsilon_n(\hbar))
\end{equation}
where $\bm F (\bm k) = \bm\nabla_k \times \bm A(\bm k)$ is the Berry curvature and $\bm A(\bm k)$ is the Berry phases in momentum space.  The intrinsic anomalous Hall effect can also be understood semi-classically by the appearance of the ``anomalous velocity''
$v_{\rm anom} = -{e\over \hbar} \bm E \times \bm F(\bm k) $ in the formula for the velocity $\bm v(k) = {1\over \hbar} {\partial \varepsilon (k) \over \partial (\bm k)} +  \bm v_{\rm anom} (\bm k)$.\cite{12}  Equation (1) is most commonly applied to the case where there is time reversal symmetry breaking by a uniform magnetization as in a ferromagnet but it is clear that $\sigma_{xy}$ can be nonzero as long as $F_z$ is nonzero in any occupied part of the Brillouin zone. Furthermore, in the case of partially filled bands, whether the band has nonzero Chern number is not necessary.

Although the results obtained here are specific to the MC state possible in the cuprates, they may be applicable to other situations.
For example the considerations may apply to the AHE discovered \cite{machida} below a phase transition to a state with no net magnetic moment or anti ferromagnetism in the pyrochlore compound Pr$_2$Ir$_2$O$_7$.

Orenstein \cite{orenstein} has also discussed  symmetry of various magneto-electric states with zero total magnetic moment per unit-cell in which a finite $\sigma_{xy}$ must exist.  Nonzero $\sigma_{xy}$ is also found for systems which break handedness but not time reversal as in a solid made up of chiral molecules \cite{LLP} and recently invoked \cite{2 refs}. Variants of the ME state with chirality in going form one plane to the other have also been recently invoked for the phenomena observed in the cuprates \cite{yakovenko}.

This paper is organized as follows. In Sec. II, we present the model Hamiltonian for the MC state in the three-orbital model for
 cuprates. Here we  also give the Haldane conditions for topological AHE in this model for a special choice of parameters of the model. The choice is necessary to get explicit analytic results but the fact that this condition exists is true for a general choice of parameters as long as the bands do not overlap. In Sec. III, numerical results are presented to show that AHE exists and we calculate its magnitude for choice of parameters consistent with those that are considered reasonable for the Cu-O model but which do not satisfy the Haldane condition.

\section{Model}

Several different translation preserving time-reversal breaking states through orbital currents are possible in the three band model of the cuprates \cite{ASV}.  Consider the MC state depicted in Fig. (\ref{Fig-AHE}). In the space of the three orbitals, $d$ for Cu, $p_x$ and $p_y$ for the two Oxygens in each unit-cell, the Hamiltonian \cite{aji-he-cmv} for such a state is
$H=\sum_{bf k} H_{\bf k}$, with
\be
\label{H}
H_{\bf k}=\left(\begin{array}{ccc}
E_d & its_x & its_y\\
-its_x & E_p & t_ps_xs_y+irc_xc_y\\
-its_y & t_ps_xs_y-irc_xc_y & E_p
\end{array}\right)
\ee
with $s_x=\sin(k_x/2)$ and $c_x=\cos(k_x/2)$, etc. Previously we considered only the case $E_d =E_p$ in which the state is topological. We consider here the more general situation $E_d \ne E_p$. In (\ref{H}), time-reversal symmetry is broken for $r \ne 0$. The Haldane condition for topological AHE restricts $r$ to be large enough that the bands do not cross. This is automatically satisfied if $E_d =E_p$. For a general $E_d - E_p$, such a condition is unlikely to be satisfied at small $r$. We will show that AHE arises even in the more general case.

In (\ref{H}), it does not matter if we subtract a constant matrix as well as make it traceless.
We can make the problem even simpler, by taking $t_p=0$. As long as $t_p$ is not too big, there is no band crossing when $t_p$ decreases to $0$. So $t_p =0$ leads to no qualitative difference in the calculated properties.
Therefore we consider the simpler Hamiltonian (where we have taken $t=1$ for convenience):

\be
H=\left(\begin{array}{ccc}
\frac23E_d & is_x & is_y\\
-is_x & -\frac13E_d & irc_xc_y\\
-is_y & -irc_xc_y & -\frac13E_d
\end{array}\right)
\label{Ham}
\ee

\subsection{Change in Chern number as $E_d$ increases from 0 to $\infty$}

For the two band model, there is a relation between Chern number and winding number which provides an analytical result for the Chern number. The 3-band model is too complicated to find a analytical formula for the Chern number directly in this manner. We will instead use an indirect way to find its Chern numbers.

When $E_d=0$, we know that the Chern numbers are $-1,\,0,\,1$ from the top to the bottom band \cite{aji-he-cmv}.
Now we show that when $E_d$ increases from 0 to $\infty$,  band degeneracy between the top and the middle band only occurs at $E_d=r$ and the Chern number of any band is non-zero only for $E_d < r$. Having a finite Chern number for a band is the general Haldane condition for a topological Anomalous Hall Effect.

The band energies are determined from
\be
&&\mbox{det}(H-\lambda I)=\lambda^3+p\lambda+q=0 \label{cubic}\\
&&p=-E_d^3/3-c_x^2c_y^2r^2-s_x^2-s_y^2\nonumber\\
&&q=\frac{E_d}{27}(-2E_d-9s_x^2-9s_y^2+18r^2c_x^2c_y^2)\nonumber
\ee
The discriminant of the above equation is
\be
&&\Delta=4p^2+27q^3=A_1+A_2+A_3+A_4\nonumber\\
&&A_1=-4r^2c_x^2c_y^2(E_d^2-c_x^2c_y^2r^2)^2\nonumber\\
&&A_2=-12r^4c_x^4c_y^4(s_x^2+s_y^2)\nonumber\\
&&A_3=-(s_x^2+s_y^2)^2[E_d^2+(s_x^2+s_y^2)]\nonumber\\
&&A_4=-4r^2c_x^2c_y^2[5E_d^2+3(s_x^2+s_y^2)](s_x^2+s_y^2)\nonumber
\ee

It is easy to see that $A_i\leq0$ for $i=1,\cdots,4$. From $A_{2,3,4}=0$, we find that $s_x=s_y=0$ or $k_x=k_y=0$.
From $A_1(k_x=0, k_y=0) =-4r^2(r^2-E_d^2)^2$, we find $E_d=r$. Therefore the only degeneracy point is located at $k_x=k_y=0$ with $E_d=r$.
In the parameter space $k_x,k_y,E_d$ this point can be thought as a monopole, as discussed earlier \cite{he-moore-v}.
When $E_d$ passes from less than $r$ to greater than $r$, there is a jump of Chern numbers for both the top and the bottom bands. But the sum of the Chern numbers of these two bands is the same as before. The change of Chern number of top band equals  the monopole charge or the Berry curvature flux through a small sphere enclosing the monopole. To see this we  expand the Hamiltonian around $k_x=k_y=0$ and $E_d=r$.
\be
H=\left(\begin{array}{ccc}
\frac23(r+\epsilon/\sqrt{2}) & i k_x/2 & i k_y/2\\
-i k_x/2 & -\frac13(r+\epsilon/\sqrt{2}) & ir\\
-i k_y/2 & -ir & -\frac13(r+\epsilon/\sqrt{2})
\end{array}\right)
\ee
To the first order in $k_x$, $k_y$ and $\epsilon$, we find the eigenstate of the top band
\be
&&\psi=\frac1{\sqrt{4R(R-\epsilon)}}\left(\begin{array}{c}
\sqrt{2}(k_x-ik_y)\\
i(\epsilon-R)\\
\epsilon-R
\end{array}
\right)
\ee
Here $R=\sqrt{\epsilon^2+k_x^2+k_y^2}$.
It is easy for this case to calculate the Berry phase and Berry curvature,
\be
A_{x,y}=-i\psi^{\dagger}\frac{\p}{\p k_{x,y}}\psi,\qquad F=\frac{\p}{\p k_x}A_y-\frac{\p}{\p k_y}A_x
\ee
Then we find that the $1/2\pi$ times the integral of the Berry curvature over the Brillouin zone is equal to $1$.
We can then show that the Chern number of the top band increases by 1 and of the middle band decreases by 1as $E_d$ goes across the degeneracy point $E_d=r$.  We will verify this below by a numerical calculation. This leads to the conclusion that for $t_p$ not too large, the Chern number of the top band is
\be
&&c_1=-1\qquad \mbox{for}\quad 0<E_d<r\\
&&c_1=0 \qquad \mbox{for}\quad  E_d>r
\ee
\noindent
So the Haldane condition for topological AHE in the simplified model is $0<E_d<r$. When this condition is not satisfied,  for $E_d>r$, the Berry phase and the Berry curvature are not identically zero, even though the top band has zero Chern number.
By taking a fixed $\epsilon>0$, we can see that $F>0$ around $k_x=k_y=0$. We must have $F<0$ in the rest of the BZ in order to cancel the positive contribution from around $k_x=k_y=0$. Then for fully filled top band $c_1=0$. To evaluate $\sigma_{xy}$ for arbitrary filling of the bands however requires a numerical calculation.

\section{Numerical result for $\sigma_{xy}$}

In the last section, we expanded the Hamiltonian around $k_x=k_y=0$, and  obtained only the formula of Berry phase and curvature around this region. In the general case, the wave function of the top band is
\be
\psi=C_N\left(\begin{array}{c}
is_y(\frac{E_d}3+E_1)-rs_xc_xc_y\\
s_xs_y-irc_xc_y(\frac{2E_d}{3}-E_1)\\
-(\frac{E_d}3+E_1)(\frac{2E_d}3-E_1)-s_x^2
\end{array}\right)
\label{wave}
\ee
Here $C_N$ is the normalization factor to make $\psi^{\dagger}\psi=1$. $E_1$ is the largest root of cubic equation Eq. (\ref{cubic}). Thus $E_1$ is an implicit function of $k_x$ and $k_y$. As $k_x$, $k_y$ vary, the order in energy of the 3 roots may change; therefore we do not have an explicit formula for $E_1$. We can only symbolically write it as
\be
E_1&=&\max_{n=0,1,2}\Big(\omega^n\sqrt[3]{-\frac{q}{2}+\sqrt{\frac{q^2}{4}+\frac{p^3}{27}}}\nonumber\\
& &\quad+\omega^{2n}\sqrt[3]{-\frac{q}{2}+\sqrt{\frac{q^2}{4}+\frac{p^3}{27}}}\Big)
\ee
Here $p$, $q$ are defined under Eq. (\ref{cubic}), and $\omega=\frac{-1+\sqrt{3}i}{2}$. We can compute $dE_1/dk_x$, etc, but the result is too lengthy to write here.
Similarly
directly plugging in Eq. (\ref{wave}) into the expressions for the Berry Phase and Berry curvature generates a very complicated result. So we pursue an alternate strategy suggested by Hatsugai  et al.\cite{Hatsugai}.

It is much easier to take the Brillouin zone for a discrete lattice and find out the eigenstates at each point numerically. Let $(i,j)$ label lattice sites.Then $k_x=\frac{2\pi i}{N}$,  $k_y=\frac{2\pi j}{N}$ for $i,j=0,1,\cdots,N$. At site $(i,j)$, numerically diagonalize $H$ and find the top band eigenvectors $\psi(i,j)$. Then we can define Berry phase as the phase difference of two eigenvector located at the neighboring sites of each links of the lattice,
\be
&&U_x(i,j)=\frac{\psi^{\dagger}(i,j)\cdot\psi(i+1,j)}{|\psi^{\dagger}(i,j)\cdot\psi(i+1,j)|},\\
&&U_y(i,j)=\frac{\psi^{\dagger}(i,j)\cdot\psi(i,j+1)}{|\psi^{\dagger}(i,j)\cdot\psi(i,j+1)|}
\ee
The $\psi(i,j)$ and $U_{x}(i,j)$ defined this way are gauge dependent and contain an arbitrary gauge factor. If we take a product of link variables $U$ around a closed loop, then the arbitrary phase factor cancels out. Therefore we can define the gauge invariant Berry curvature for each plaquette as
\be
F(i,j)&=&-i\ln\Big[U_x(i,j)U_y(i+1,y)\nonumber\\
& &\qquad\times U_x^{-1}(i,j+1)U_y^{-1}(i,j)\Big]
\label{F}
\ee
Here $F$ is just the phase angle of the quantity inside the bracket; as a convention we assume $-\pi<F<\pi$.
Then the Chern number is
\be
c=\frac1{2\pi}\sum_{i,j}F(i,j)
\ee
All the above are standard definitions of vector field and field strength in lattice gauge field theory. Hatsugai et al. \cite{Hatsugai} point out that this method provides a very efficient way to compute Chern number numericaly \cite{Hatsugai}. Even for a very small lattice such as 4 by 4, it  generates the correct Chern number. The following results are obtained by this method.

\begin{figure}
\centerline{\includegraphics[width=0.4\textwidth]{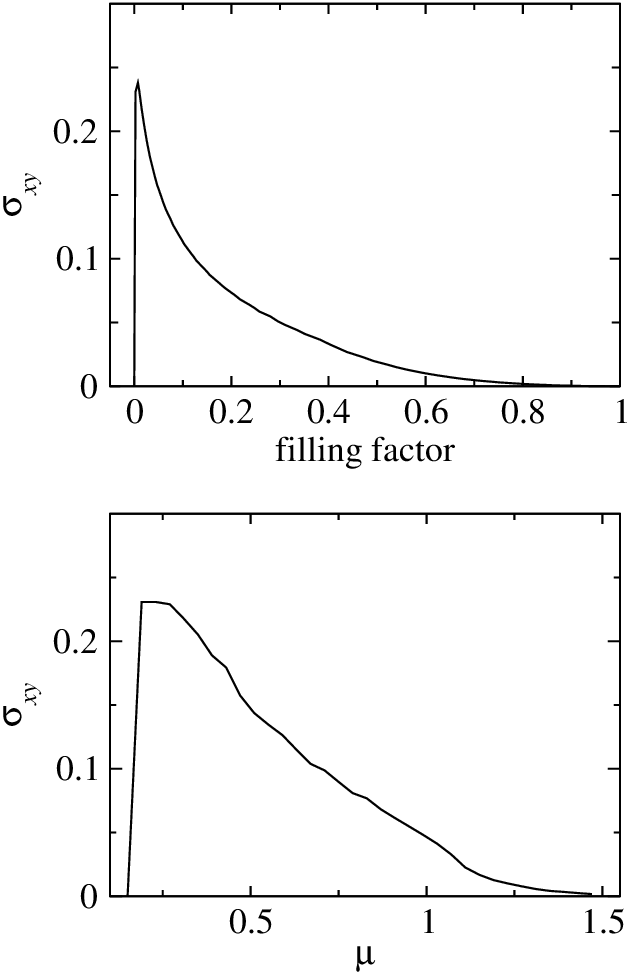}}
\caption{$\sigma_{xy}$ as a function of filling and $\mu$.}
\label{sig}
\end{figure}

 The Hall conductance is the integration of Berry curvature up to the Fermi surface. Now we show the numerical results of the partially filled band conductance $\sigma_{xy}$ in figure (\ref{sig}). In Figure (\ref{sig}), we take $t=1$, $t_p=0.5$ $r=0.1$ and $E_d=0.15$. In the left panel, we plot $\sigma_{xy}$ as a function of filling factor. In the right panel we plot $\sigma_{xy}$ as a function of the chemical potential $\mu$.  Since the energy minima of the top band is located at $k_x=k_y=0$, the fermions  first fill the region around $k_x=k_y=0$. In this region the Berry curvature is positive; therefore $\sigma_{xy}$ is positive at first. When the fermi energy or the filing factor increases, we integrate an increasingly larger area such that we begin to get a negative contributions from the points far away from $k_x=k_y$. When the filling approaches 1, $\sigma_{xy}$ gradually approaches 0. Thus for these general parameters which do not satisfy the Haldane restrictions, AHE exists only for a partial filling of a band.

\begin{figure}
{\includegraphics[width=0.4\textwidth]{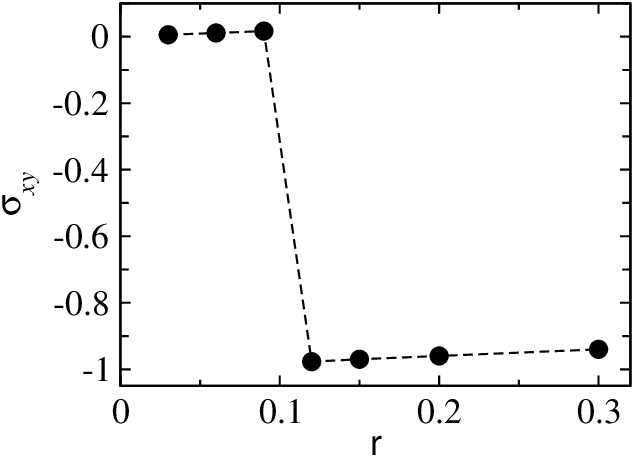}}
\caption{Half filling $\sigma_{xy}$ as a function of $r$.}
\label{sig-r}
\end{figure}

In the above discussion  $r$ is fixed and $E_d$ is changed. This is equivalent to fixing $E_d$ and sweeping $r$. In figure \ref{sig-r}, we take $t=1$, $t_p=0.5$ and $E_d=0.1$, and plot $\sigma_{xy}$ at half-filling as a function of $r$. From Eq. (\ref{F}), we can see that around BZ center ($k_x=k_y=0$), $F>0$ for $E_d>r$ and $F<0$ for $E_d<r$. In both cases, the band bottom is at BZ center. For half filling, the filled region is roughly a circle enclosing the BZ center, thus $\sigma_{xy}>0$ for $E_d>r$ and $\sigma_{xy}<0$ for $E_d<r$. This is why there is a jump when $r$ passing through $E_d$.
For $r<E_d$, at half filling, $\sigma_{xy}$ is already quite close to the Chern number 0, thus it is a small positive number. One the other hand, for $r>E_d$, $\sigma_{xy}$ is also quite close to the Chern number -1, thus it is a negative number slightly above -1.

\section{Conclusions}

We have shown that partially filled bands, one can have a finite $\sigma_{xy}$ for states with broken time-reversal symmetry but with zero net flux per unit-cell in conditions more general than required for Haldane type topological states. In Haldane type states, the $\sigma_{xy}$ for each filled band band is zero or a positive or negative integer in units of $e^2/h$ such that the sum over states of all the bands is zero. This is achieved for parameters in the effective one-particle Hamiltonian such that there is a band-crossing such that one may speak of a monopole in the Brillouin zone. When the condition are such that time-reversal is broken but no topological object such as a monopole exists, one may still have a finite hall effect for a partially filled band while it is zero when summed over states of each band. The physical point is that time-reversal with net zero moment per unit-cell has finite current carried by each ${\bf k}$ state. If the symmetry is such that a given ${\bf k}$ state does not have a degenerate partner under reflection or inversion carrying current in the opposite direction, there is a finite $\sigma_{xy}$.

One finds on comparing the results presented here in Fig. (\ref{sig}, \ref{sig-r}) for near half-filling and $r \approx 0.1$, which is the relevant range with results for similar parameters in ref. (\onlinecite{aji-he-cmv}) that they have the same order of magnitude. To get Kerr effect of the (small) magnitude observed in experiments requires an additional small parameter, the reflection breaking lattice distortion, as discussed previously \cite{aji-he-cmv} of $O(10^{-3})$. As already stated in  Ref. \onlinecite{aji-he-cmv}, we do not know of any direct determination of this number. There are other remarkable features of the experiments which were explained in  Ref. \onlinecite{aji-he-cmv}. These are the independence of the effect on the presence of domains, lack of history dependence on cooling and warming and getting the same sign of the effect from shining light on opposite surfaces. These properties only depend on the magneto-chiral nature of the induced order and remain true for the results in this paper. It is also worth reminding that experiments \cite{kerr-expt} show the occurrence of the Kerr effect coincident with the ocurrence  of the pseudo-gap in a family of cuprates and occurring below the pseudo-gap temperature but heading towards $T \to 0$ as a function of doping at the same doping where the pseudo-gap  tends to go to zero. This is consistent with the idea which is the basis of this as well as the previous work \cite{aji-he-cmv} that time-reversal breaking ( as observed by neutron-scattering at the pseudo-gap temperature quite generally)  as well as lattice distortion is necessary to have the observed Kerr effect.

The results on the AHE presented here are quite general and require only a system with partially filled bands and a magneto-chiral symmetry. The strange AHE observed in Ref.(\onlinecite{machida}) may be an example, besides the cuprates. To test the idea microscopically requires polarized neutron scattering experiments as in Refs. (\onlinecite{neutrons}) or other experiments which observe the requisite Magnetic Bragg spots.

{\it Acknowledgements} P.A. Lee acknowledges the support by NSF grant number DMR 1104498.
and C.M. Varma acknowledges support by NSF grant number DMR 1206298.

\end{document}